\documentclass[aps,showpacs,showkeys,superscriptaddress,11pt]{revtex4}

\usepackage{graphicx,epic,eepic,epsfig,amsmath,latexsym,amssymb,verbatim}

\usepackage{theorem}
\newtheorem{definition}{Definition}[section]

\newtheorem{lemma}[definition]{Lemma}

\newtheorem{theorem}[definition]{Theorem}

\def\squareforqed{\hbox{\rlap{$\sqcap$}$\sqcup$}}
\def\qed{\ifmmode\squareforqed\else{\unskip\nobreak\hfil
\penalty50\hskip1em\null\nobreak\hfil\squareforqed
\parfillskip=0pt\finalhyphendemerits=0\endgraf}\fi}
\def\endenv{\ifmmode\;\else{\unskip\nobreak\hfil
\penalty50\hskip1em\null\nobreak\hfil\;
\parfillskip=0pt\finalhyphendemerits=0\endgraf}\fi}
\newenvironment{proof}{\noindent \textbf{{Proof~} }}{\qed}

\newcommand{\exampleTitle}[1]{\textbf{(#1)}}

\newcommand{\proofComment}[1]{\exampleTitle{#1}}

\mathchardef\ordinarycolon\mathcode`\:
\mathcode`\:=\string"8000
\def\vcentcolon{\mathrel{\mathop\ordinarycolon}}
\begingroup \catcode`\:=\active
  \lowercase{\endgroup
  \let :\vcentcolon
  }

\newcommand{\nc}{\newcommand}
\nc{\rnc}{\renewcommand}
\nc{\beq}{\begin{equation}}
\nc{\eeq}{{\end{equation}}}
\nc{\beqa}{\begin{eqnarray}}
\nc{\eeqa}{\end{eqnarray}}
\nc{\lbar}[1]{\overline{#1}}
\nc{\bra}[1]{\langle#1|}
\nc{\ket}[1]{|#1\rangle}
\nc{\ketbra}[2]{|#1\rangle\!\langle#2|}
\nc{\braket}[2]{\langle#1|#2\rangle}
\nc{\proj}[1]{| #1\rangle\!\langle #1 |}
\nc{\avg}[1]{\langle#1\rangle}
\rnc{\max}{\operatorname{max}}
\nc{\Rank}{\operatorname{Rank}}
\nc{\smfrac}[2]{\mbox{$\frac{#1}{#2}$}}
\nc{\Tr}{\operatorname{Tr}}
\nc{\ox}{\otimes}
\nc{\dg}{\dagger}
\nc{\dn}{\downarrow}
\nc{\cA}{{\cal A}}
\nc{\cB}{{\cal B}}
\nc{\cC}{{\cal C}}
\nc{\cD}{{\cal D}}
\nc{\cE}{{\cal E}}
\nc{\cF}{{\cal F}}
\nc{\cG}{{\cal G}}
\nc{\cH}{{\cal H}}
\nc{\cI}{{\cal I}}
\nc{\cJ}{{\cal J}}
\nc{\cK}{{\cal K}}
\nc{\cL}{{\cal L}}
\nc{\cM}{{\cal M}}
\nc{\cO}{{\cal O}}
\nc{\cP}{{\cal P}}
\nc{\cR}{{\cal R}}
\nc{\cS}{{\cal S}}
\nc{\cT}{{\cal T}}
\nc{\cX}{{\cal X}}
\nc{\cZ}{{\cal Z}}
\nc{\csupp}{{\operatorname{csupp}}}
\nc{\qsupp}{{\operatorname{qsupp}}}
\nc{\var}{\operatorname{var}}
\nc{\rar}{\rightarrow}
\nc{\lrar}{\longrightarrow}
\nc{\polylog}{\operatorname{polylog}}

\def\a{\alpha}

\def\g{\gamma}
\def\d{\delta}
\def\e{\epsilon}

\def\l{\lambda}

\def\r{\rho}
\def\s{\sigma}

\def\ph{\varphi}

\def\ps{\psi}

\def\D{\Delta}

\def\L{\Lambda}

\def\Ph{\Phi}

\nc{\RR}{{{\mathbb R}}}
\nc{\CC}{{{\mathbb C}}}
\nc{\FF}{{{\mathbb F}}}
 \nc{\NN}{{{\mathbb N}}}
\nc{\ZZ}{{{\mathbb Z}}}
\nc{\PP}{{{\mathbb P}}}
\nc{\QQ}{{{\mathbb Q}}}
\nc{\UU}{{{\mathbb U}}}
\nc{\EE}{{{\mathbb E}}}
\nc{\id}{{\mathbb I}}

\nc{\be}{\begin{equation}}
\nc{\ee}{{\end{equation}}}
\nc{\bea}{\begin{eqnarray}}
\nc{\eea}{\end{eqnarray}}
\nc{\<}{\langle}
\rnc{\>}{\rangle}
\nc{\Hom}[2]{\mbox{Hom}(\CC^{#1},\CC^{#2})}
\nc{\rU}{\mbox{U}}
\def\lpmm{ \left(\rule{0pt}{1.8ex}\right. \! }
\def\rpmm{ \! \left.\rule{0pt}{1.8ex}\right) }
\def\lbm{ \left[\rule{0pt}{2.1ex}\right. }
\def\rbm{ \left.\rule{0pt}{2.1ex}\right] }
\def\lpm{ \left(\rule{0pt}{2.1ex}\right. \!}
\def\rpm{ \!\left.\rule{0pt}{2.1ex}\right) }
\def\lbL{ \left[\rule{0pt}{2.4ex}\right. \!}
\def\rbL{ \!\left.\rule{0pt}{2.4ex}\right] }
\def\lpL{ \left(\rule{0pt}{2.4ex}\right.\!\!}
\def\rpL{ \!\! \left.\rule{0pt}{2.4ex}\right)}
\def\lpH{ \left(\rule{0pt}{3.0ex}\right.\!\!}
\def\rpH{ \!\! \left.\rule{0pt}{3.0ex}\right)}
\def\llH{ \left|\rule{0pt}{3.0ex}\right.\!}
\def\rlH{ \!\left.\rule{0pt}{3.0ex}\right|}
\def\llL{ \left|\rule{0pt}{2.4ex}\right.\!}
\def\rlL{ \!\left.\rule{0pt}{2.4ex}\right|}
\def\llm{ \left|\rule{0pt}{2.1ex}\right.\!}
\def\rlm{ \!\left.\rule{0pt}{2.1ex}\right|}
\nc{\ob}[1]{#1}

\nc{\figurePQC}{
	\begin{figure}
	\begin{center}
	\setlength{\unitlength}{0.00043333in}
{\renewcommand{\dashlinestretch}{30}
\begin{picture}(8659,3162)(0,-10)
\put(6675,2337){\blacken\ellipse{300}{300}}
\put(1875,2337){\blacken\ellipse{300}{300}}
\path(225,2412)(3450,2412)
\path(225,2337)(3450,2337)
\path(5100,2412)(8325,2412)
\path(5100,2337)(8325,2337)
\path(225,537)(1350,537)
\path(2400,537)(6150,537)
\path(7200,537)(8325,537)
\path(1875,2337)(1875,1212)
\path(6675,2337)(6675,1212)
\whiten\path(1350,1212)(2400,1212)(2400,12)
	(1350,12)(1350,1212)
\path(1350,1212)(2400,1212)(2400,12)
	(1350,12)(1350,1212)
\whiten\path(6150,1212)(7200,1212)(7200,12)
	(6150,12)(6150,1212)
\path(6150,1212)(7200,1212)(7200,12)
	(6150,12)(6150,1212)
\dottedline{45}(3600,2367)(4950,2367)
\put(1750,537){$U_j$}
\put(6525,537){$U_j^\dagger$}
\put(1650,3012){Alice}
\put(6525,3012){Bob}
\put(375,2562){$j$}
\put(5250,2562){$j$}
\put(0,537){$\varphi$}
\put(3900,687){$R(\varphi)$}
\put(8475,537){$\varphi$}
\end{picture}
}
	\end{center}
	\caption{A private quantum channel built on
	the randomization map $R$. Alice and Bob share knowledge of the secret
	key $j$, using it to encrypt and decrypt the state. Because an 
	eavesdropper does not have access to $j$, her view is 
	$R(\ph) \approx \id / d$. If $\ph$ is a $d$-dimensional quantum 
	state, then the key length need only be $\log d + o(\log d)$.}
	\label{fig:pqc}
	\end{figure}
}

\nc{\figureHiding}{
	\begin{figure}
	\begin{center}
	\setlength{\unitlength}{0.00053333in}
\begingroup\makeatletter\ifx\SetFigFont\undefined%
\gdef\SetFigFont#1#2#3#4#5{%
  \reset@font\fontsize{#1}{#2pt}%
  \fontfamily{#3}\fontseries{#4}\fontshape{#5}%
  \selectfont}%
\fi\endgroup%
{\renewcommand{\dashlinestretch}{30}
\begin{picture}(11858,3253)(0,-10)
\put(4424.000,2491.750){\arc{292.055}{0.8715}{2.2701}}
\put(4423.979,2730.023){\arc{358.401}{3.1917}{6.3280}}
\put(4424.000,2986.308){\arc{666.859}{1.0041}{2.1375}}
\put(4424.000,2491.750){\arc{292.055}{0.8715}{2.2701}}
\put(4424.000,2508.750){\arc{292.055}{0.8715}{2.2701}}
\texture{44555555 55aaaaaa aa555555 55aaaaaa aa555555 55aaaaaa aa555555 55aaaaaa 
	aa555555 55aaaaaa aa555555 55aaaaaa aa555555 55aaaaaa aa555555 55aaaaaa 
	aa555555 55aaaaaa aa555555 55aaaaaa aa555555 55aaaaaa aa555555 55aaaaaa 
	aa555555 55aaaaaa aa555555 55aaaaaa aa555555 55aaaaaa aa555555 55aaaaaa }
\put(4424,2730){\shade\ellipse{358}{118}}
\put(4424,2730){\ellipse{358}{118}}
\path(4245,2722)(4330,2380)
\path(4603,2722)(4518,2380)
\path(4535,2653)(4484,2363)(4501,2380)
	(4569,2670)(4535,2653)
\put(2475,2644){\blacken\ellipse{300}{300}}
\put(2475,2644){\ellipse{300}{300}}
\thicklines
\put(9556,1476){\ellipse{4574}{2924}}
\thinlines
\put(8699,2189){\ellipse{1050}{900}}
\put(9656,2143){\ellipse{1050}{900}}
\put(9131,1543){\ellipse{1050}{900}}
\put(10321,1567){\ellipse{1050}{900}}
\put(9796,742){\ellipse{1050}{900}}
\put(7996,1267){\ellipse{1050}{900}}
\put(8746,742){\ellipse{1050}{900}}
\put(10846,892){\ellipse{1050}{900}}
\path(825,2719)(4050,2719)
\path(825,2644)(4050,2644)
\whiten\path(1950,1894)(3000,1894)(3000,19)
	(1950,19)(1950,1894)
\path(1950,1894)(3000,1894)(3000,19)
	(1950,19)(1950,1894)
\path(825,919)(1950,919)
\path(3000,1519)(5100,1519)
\path(3000,394)(5100,394)
\path(5175,1744)(5325,1744)(5325,169)(5175,169)
\path(2475,2494)(2475,1894)
\put(2400,844){$U_j$}
\put(975,2869){$j$}
\put(4000,1669){Alice: ${\cal H}_A$}
\put(4000,94){Bob: ${\cal H}_B$}
\put(450,1044){$|\varphi\rangle\in S$}
\put(5475,919){$R(\varphi)$}
\put(8925,1444){$U_1 S$}
\put(8425,2094){$U_2 S$}
\put(9525,2094){$U_3 S$}
\put(10050,1519){$U_4 S$}
\put(10625,844){$U_5 S$}
\put(9600,694){$U_6 S$}
\put(8625,694){$U_7 S$}
\put(7800,1219){$U_8 S$}
\put(10325,2144){${\cal H}_A \otimes {\cal H}_B$}
\put(0,3094){(a)}
\put(6525,3094){(b)}
\end{picture}
}
	\end{center}
	\caption{Quantum data hiding. (a) depicts the encoding procedure. 
	A random $U_j$ is applied to the state $\ket{\ph}$ drawn from 
	subspace $S$. The output, $R(\ph)$, is almost indistinguishable 
	from the maximally mixed state using LOCC alone. (b) The different 
	subspaces $\{U_j S\}$ have very small overlaps, however, so a 
	collective operation on $\cH_A \ox \cH_B$ can be used to
	distinguish them without causing much distortion to the encoded states
	$U_j \ket{\ph}$.}
	\label{fig:hiding}
	\end{figure}
}

\begin{document}

\title{{\Large Randomizing quantum states:} \\ Constructions and applications}

\author{Patrick Hayden}
\email{patrick@cs.caltech.edu}
\affiliation{Institute for Quantum Information, Caltech 107--81,
    Pasadena, CA 91125, USA}
\affiliation{Mathematical Sciences Research Institute, 1000 Centennial Drive, 
    Berkeley, CA 94720, USA}
\author{Debbie~Leung}
\email{wcleung@cs.caltech.edu}
\affiliation{Institute for Quantum Information, Caltech 107--81,
    Pasadena, CA 91125, USA}
\affiliation{Mathematical Sciences Research Institute, 1000 Centennial Drive, 
    Berkeley, CA 94720, USA}
\author{Peter W. Shor}
\email{shor@research.att.com}
\affiliation{AT \& T Labs Research, Florham Park, NJ 07922, USA}
\author{Andreas Winter}
\email{winter@cs.bris.ac.uk}
\affiliation{Department of Computer Science, University of Bristol,\\
Merchant Venturers Building, Woodland Road, Bristol BS8 1UB, United Kingdom}
\affiliation{Mathematical Sciences Research Institute, 1000 Centennial Drive, 
    Berkeley, CA 94720, USA}

\date{\today}

\begin{abstract}
The construction of a perfectly secure private quantum channel in
dimension $d$ is known to require $2 \log d$ shared random key bits
between the sender and receiver. We show that if only near-perfect
security is required, the size of the key can be reduced by a factor
of two. More specifically, we show that there exists a set of 
roughly $d \log d$ unitary operators whose average effect on every input
pure state is almost perfectly randomizing, as compared to the $d^2$
operators required to randomize perfectly.  Aside from the private
quantum channel, variations of this construction can be applied to
many other tasks in quantum information processing. We show, for
instance, that it can be used to construct LOCC data hiding schemes
for bits and qubits that are much more efficient than any others
known, allowing roughly $\log d$ qubits to be hidden in $2 \log d$
qubits. The method can also be used to exhibit the existence of
quantum states with locked classical correlations, an arbitrarily
large amplification of the correlation being accomplished by sending a
negligibly small classical key. Our construction also provides the
basic building block for a method of remotely preparing arbitrary 
$d$-dimensional pure
quantum states using approximately $\log d$ bits of communication and
$\log d$ ebits of entanglement.
\end{abstract}



\maketitle

\section{Introduction} \label{sec:intro}
In this paper we revisit the question of finding the minimal resources
required to randomize a quantum state. This problem has previously
been investigated in several variations, always with the conclusion
that in order to randomize or, more generally, encrypt a quantum state
of $l$ qubits, $2l$ classical bits of random key are
required~\cite{BLS99,BR00,AMTW00}.  This factor of $2$ represents a
familiar and even welcome phenomenon in quantum information theory;
the reason for its appearance is intimately connected to the existence
of superdense coding~\cite{AMTW00,BW92}.  All this previous work,
however, considered only the task of \emph{perfectly} encrypting
quantum states.  Here we focus on the task of \emph{approximately}
encrypting quantum states, allowing a negligible but non-zero amount
of information to remain available to an eavesdropper. In sharp
contrast to the exact case, in this setting we find that the factor of
$2$ disappears entirely: an $l$-qubit quantum state can be
approximately encrypted using $l + o(l)$ bits of random key.

Our encryption or, more specifically, randomization scheme also 
exposes a previously
unobserved difference between classical and quantum correlations:
classical correlations are always effectively destroyed by a local
randomization procedure whereas quantum correlations need not be.
Therefore, any correlation that survives local randomization must be 
``nonlocal'' and, hence, quantum mechanical in nature.  
This basic insight provides the intuition behind a new
scheme for data hiding in a bipartite system~\cite{DLT02,EW02,DHT02}.  
The encoding is an approximate randomization
procedure applied collectively to the two $l$-qubit shares of a $2l$-qubit
system, calibrated to eliminate all correlations that can 
be detected by local operations and classical communication (LOCC).  
The failure of the encoding to destroy quantum correlations is striking:
there exists a collective decoding operation that recovers all the states
on an $[l - o(l)]$-qubit subsystem of the input.
In other words, randomization can be used to construct
schemes for LOCC hiding of roughly $l$ qubits in a $2l$-qubit quantum
state. This construction is far more efficient than any previously
known for hiding qubits or even classical bits.

Another variation on our basic construction can be used to find quantum
states whose classical correlations are large but \emph{locked}. Roughly
speaking, this means that local measurements can yield classical
data with only a small amount of correlation but that local operations
supplemented with a
small amount of communication can yield a disproportionately
large amount of correlation. The existence of
such states was demonstrated in Ref.~\cite{DHLST03} but some central
questions about the range of possible effects were left open.
In particular, the authors defined two figures of merit, one measuring
the (reciprocal of) \emph{amplification} of the correlation and the other
the ratio of the amount of communication to the amount of unlocked
correlation. We give the first demonstration that both quantities can
go to zero simultaneously. 
While seemingly esoteric, the existence of this
phenomenon has important implications for the definition of security in
quantum cryptographic scenarios; in particular, it establishes the
potentially enormous volatility of accessible information.
There is also an alternative interpretation of this result: we 
prove that random observables typically obey extremely strong
entropic uncertainty relations.

One final application of the ability to randomize an $l$-qubit state
using only $l + o(l)$ random key bits has a sufficiently different
character that we present it in a separate paper~\cite{BHLSW03}.
Insofar as teleportation~\cite{BBCJPW93}, or more generally
\emph{remote state preparation}~\cite{Lo99}, can be interpreted as a
method for encrypting quantum states~\cite{Leung00,LS02}, our
approximate encryption procedure should give rise to an approximate
remote state preparation method consuming only half as much
communication as teleportation.  We report in \cite{BHLSW03} how
our results on randomization provide the basic building block for a
protocol capable of sending an arbitrary pure $l$-qubit quantum
state using only $l$ ebits and $l+o(l)$ bits of classical
communication.

The rest of the paper is structured as follows. Section \ref{sec:Haar}
describes our results on the private quantum channel, which serve as a
prototype for the rest of the paper. Section \ref{sec:characterize}
then studies approximately randomizing maps by characterizing their
effect on classical and quantum correlations. These observations are
then put to use in slightly modified form in section \ref{sec:dataHiding},
which describes our quantum data hiding protocol and proves both
its correctness and security. Section \ref{sec:locking} then formalizes
the idea of locking classical correlations and describes our contribution.
We also include an appendix establishing some results on randomization
procedures using subsets of the unitary group, such as the Pauli matrices.

We use the following conventions throughout the paper.  
$\log$ and $\exp$ are always taken base $2$. Unless
otherwise stated, a ``state'' can be pure or mixed.  The symbol for a state
(such as $\ph$ or $\rho$) also denotes its density matrix.  ``Part of
an entangled state'' refers to a (mixed) state whose purification is
accessible to some of the parties.  We will make an explicit
indication when referring to a {\em pure} state. The density operator
$\proj{\ph}$ of the pure state $\ket{\ph}$ will frequently be written simply
as $\ph$. $\cB(\CC^d)$ will be used to denote the set
of linear operators from $\CC^d$ to itself and
$\rU(d) \subset \cB(\CC^d)$ the 
unitary group on $\CC^d$.

Finally, a word of warning about the cryptographic interpretation of our
results. When we say that a scheme for approximate encryption or quantum 
data hiding is secure for mixed states, it should be assumed that the 
purifications of those mixed states are inaccessible to all parties 
considered. Indeed, the possibility that purification-inaccessible
security criteria can hold even as purification-accessible criteria
fail is essentially a quantum mechanical re-statement of a familiar
cryptographic observation: approximate security is sometimes much 
easier to achieve than perfect security. We exploit the gap throughout
this paper.

\section{An approximate private quantum channel} \label{sec:Haar}
We consider an insecure one-way quantum channel between two parties
Alice and Bob that is noiseless in the absence of eavesdropping.  This
channel can be made secure against eavesdropping if Alice and Bob are
allowed the extra resource of shared random secret key bits. For
instance, one can encrypt a state $\ket{\ph} \in \CC^d$ using a
secret key of length $2 \log d$ as follows~\cite{BR00,AMTW00}. Fix a
basis $\{\ket{\ob{1}},\ldots,\ket{\ob{d}}\}$ for $\CC^d$ and let
\begin{equation}
X \ket{\ob{j}} = \ket{\ob{(j+1) \bmod d}} 
\quad \mbox{and} \quad
Z \ket{\ob{j}} = e^{2\pi i j / d} \ket{\ob{j}}.
\end{equation}
It's straightforward to verify that
\begin{equation} \label{eqn:perfectRandom}
\frac{1}{d^2} \sum_{j=1}^{d} \sum_{k=1}^d X^j Z^k \ph {Z^k}^\dg {X^j}^\dg
= \frac{\id}{d}
\end{equation}
for all states $\ph$. 
If $j$ and $k$ are selected using a shared secret
key, Alice can encrypt the state using the unitary operation $X^j Z^k$ and
Bob can decrypt by applying $Z^{-k} X^{-j}$. By Eq.~(\ref{eqn:perfectRandom}),
the view of an eavesdropper without access to $j$ and $k$ is 
$\id /d$, which is independent of the input state $\ph$.
This structure, consisting of a set of encoding maps and decoding maps
indexed by key values, such that the average encoded state is independent
of the input, is known as a \emph{private quantum channel} for $\CC^d$.
(See Ref.~\cite{AMTW00} for a formal definition.)
In fact, if perfect
security is required, it can be shown that the secret key \emph{must}
have length at least $2 \log d$~\cite{BLS99,BR00,AMTW00,LS02}.
Here we relax the security criterion.
\begin{definition} \label{approxrand} A completely positive, trace-preserving
(CPTP) map $R\! :\! \cB(\CC^d) \! \rightarrow \! \cB(\CC^d)$ is
\emph{$\e$-randomizing} if, for all states $\ph$, 
\begin{equation} \label{eqn:eRandomizing}
\left\| R(\ph) - \frac{\id}{d} \right\|_\infty \leq \frac{\e}{d} \,.
\end{equation}
\end{definition}
In the above, $\|\cdot\|_\infty$ is the operator norm, 
so Eq.~(\ref{eqn:eRandomizing}) 
is equivalent to all the eigenvalues of $R(\ph)$ lying in the interval 
$[(1-\e)/d,(1+\e)/d]$.
By convexity of the norm, it suffices to check the condition for
all pure states, a fact we will use repeatedly. 
In quantum information, distinguishability is frequently measured
using the trace norm $\|\cdot\|_1 = \Tr|\cdot|$, the analogue of 
variation distance in probability theory. Note that 
Eq.~(\ref{eqn:eRandomizing}) automatically implies the
weaker estimate $\| R(\ph) - \id / d \|_1 \leq \e$. The main result of
this section is
\begin{theorem} \label{thm:bigRandom}
For all $\e>0$ and sufficiently large $d$ ($> \! \smfrac{10}{\e}$),
there exists a choice of unitaries in $\rU(d)$, $\{U_j:1\leq j \leq n
\}$ with $n = 134 d (\log d) / \e^2$ such that the map
\begin{equation}      
R(\ph) = \frac{1}{n} \sum_{j=1}^n U_j \ph U_j^\dg
\label{eq:defr}
\end{equation}
on $\cB(\CC^d)$ is $\e$-randomizing.  
\end{theorem}
As illustrated in figure \ref{fig:pqc}, by having Alice and Bob select
$j$ using a shared secret key, this map $R$ can be used to build a
private quantum channel with key length
$\log n = \log d + \log \log d + \log(1/\e^2) + 8$, albeit one that is not
perfectly secure, only nearly so. 
The view of an eavesdropper without access to the key, 
given a particular input state $\ph$, is precisely $R(\ph)$. Definition
\ref{approxrand}, therefore, doubles as a definition of security.
For any distribution of states $\{p_i,\ph_i\}$ supported on $\CC^d$ we
can bound the mutual information accessible to an eavesdropper who
performs measurements on the encrypted states $R(\ph_i)$. This 
accessible information is bounded above by the Holevo quantity~\cite{Holevo73}
\begin{eqnarray}
\chi 
&=& S\left( \sum_i p_i R(\ph_i) \right) - \sum_i p_i S( R(\ph_i) ) 
	\\
&\leq& \log d - \sum_i p_i S\big( R(\ph_i) \big)
	\\
&\leq& \log(1+\e)	
\leq \e / (\ln 2).
\end{eqnarray}
The second inequality is true because the definition of 
$\e$-randomizing maps implies that $R(\ph_i) \leq (1+\e)\id / d$,
which allows for an application of the monotonicity of $\log$.
%
In particular, for all $\a>0$, choosing 
$\e = (\log d)^{-\a}$ implies that
$\chi 
\xrightarrow{d \rar \infty} 0$.
In this case, $n = 134 \, d \, (\log d)^{(1+2\a)}$, so the key size can
be taken to be $\log d + (1+2\a) \log \log d + 8$.  
Alternatively, one can choose an arbitrarily small $\a>0$, and let 
$\e = d^{-\a}$.  Then, $\chi \leq d^{-\a} / (\ln 2)$ which is
exponentially decaying in the number of qubits to be encrypted, 
$\log d$.
This comes at the cost of a slight increase in the asymptotic key
length: $\log n = (1+2\a) \log d + \log \log d + 8$. 
Moreover, as is 
common with probabilistic existence proofs, it is possible to ensure
that the overwhelming majority of random choices succeed at only
minor additional cost; in this case, $\log n$ would need to be increased by a
constant number of bits for any fixed probability of success. 
Analogous statements hold for our other 
constructions later in the paper.
 
\figurePQC

The proof of theorem \ref{thm:bigRandom} is based on a large deviation
estimate, lemma \ref{lem:conc}, and discretization via a net
construction, lemma \ref{lem:net}, both of which will be re-used in
other applications later in the paper.  
The large deviation estimate, in turn, is based on Cram\'{e}r's theorem
(see Ref.~\cite{DZ93}, for example, for a detailed exposition), 
which states that for independent,
identically distributed (i.i.d.) real-valued
random variables, $X,X_1,X_2,\ldots,X_n$, 
\begin{eqnarray}
\Pr\lpH \frac{1}{n} \sum_{j=1}^n X_j \geq a \rpH
&\leq& \exp\left( -n \frac{1}{\ln 2} \inf_{x \geq a} \L^*(x) \right)
\quad \mbox{and} 
\label{eq:cramerlarger}
\\
\Pr\lpH \frac{1}{n} \sum_{j=1}^n X_j \leq a \rpH 
&\leq& \exp\left( -n \frac{1}{\ln 2} \inf_{x \leq a} \L^*(x) \right),
\label{eq:cramersmaller}
\end{eqnarray}
where 
\begin{equation}
\L^*(x) = \sup_{\l \in \RR} \; \lbm \l x - \ln \EE e^{\l X} \rbm. 
\end{equation}
$\L^*(x)$ is known as the \emph{rate function} and 
$\EE e^{\l X}$ the \emph{moment generating function}. 
(In fact, the harder part of Cram\'{e}r's theorem deals with the 
optimality of the rate function. We only need bounds (\ref{eq:cramerlarger})
and (\ref{eq:cramersmaller}) here, which are surprisingly easy to prove: they
require only the Bernstein trick and Markov's inequality.) 

\begin{lemma} \label{lem:conc}
Let $\ph$ be a pure state, $P$ a rank $p$ projector and let
$(U_j)_{j\geq 1}$ be an i.i.d.
sequence of $\rU(d)$-valued random variables, distributed according to
the Haar measure. There exists a constant $C$ ($C \geq (6 \ln
2)^{-1}$) such that if $0 < \e < 1$,
\begin{equation}
\Pr\left( \; \left| \frac{1}{n} \sum_{j=1}^n
    \Tr(U_j \ph U_j^\dg P) - \frac{p}{d} \, \right|
    \geq \frac{\e p}{d} \right)
\leq
2 \exp\left( - C np \e^2 \right).
\end{equation}
\end{lemma}
\begin{proof}
Since the Haar measure is left and right invariant, we may assume that
$\ph = \proj{\ob{1}}$ and $P = \sum_{i=1}^p \proj{\ob{i}}$ for some
fixed orthonormal basis $\{\ket{\ob{i}}\}$.  Let $\ket{g_j} =
\sum_{i=1}^d g_{ij} \ket{\ob{i}}$, where the i.i.d.~complex random
variables $g_{ij} \sim N_\CC(0,1)$. (That is, the real and imaginary parts 
of $g_{ij}$ are independent gaussian random variables with mean $0$ and 
variance $1/2$.)  The distribution of $\ket{g_j}$ is the same as the
distribution for $\| g_j \|_2 \, U_j \ket{\ob{1}}$.  
For a fixed $j$, let $U = U_j$ and
$\ket{g}=\ket{g_j}$.  The convexity of $\exp$ implies that
\begin{eqnarray}
\EE_g \exp \left( \frac{\l}{d} \sum_{i=1}^p | \braket{\ob{i}}{g} |^2 \right)
&=& \EE_U \EE_g \exp\left( \frac{\l \| g \|_2^2}{d}
    \sum_{i=1}^p | \< \ob{i}| U |\ob{1}\> |^2 \right) \\
&\geq& \EE_U \exp\left( \EE_g \frac{\l \| g \|_2^2}{d}
    \sum_{i=1}^p | \<\ob{i}| U |\ob{1}\> |^2 \right)  \\
&=& \EE_U \exp \lpL \l \sum_{i=1}^p |\< \ob{i}| U |\ob{1}\> |^2 \rpL \\
&=& \EE_U \exp \lpm \l \Tr(U \ph U^\dg P) \rpm \,.
\end{eqnarray}
This inequality between moment generating functions establishes, via
Cram\'{e}r's theorem, that
the rate function $\L_U^*$ for the random variable $\Tr(U\ph U^\dg P)$
and the rate function $\L_p^*$ for 
$\smfrac{1}{d} \sum_{i=1}^p |\braket{i}{g}|^2$ are related by the
inequality $\L_p^*(x) \leq \L_U^*(x)$. It follows from the definitions
that $\L_p^*(px/d)$, in turn, is equal
to $p \L_g^*(x)$, where $\L_g^*$ is the rate function for 
$|g_{ij}|^2$. 
Therefore,
\begin{equation}
\Pr\left( \frac{1}{n} \sum_{j=1}^n
    \Tr(U_j \ph U_j^\dg P) - \frac{p}{d} 
    \geq \frac{\e p}{d} \right)
\leq
\exp\left( -np \frac{1}{\ln 2} \inf_{x \geq \e} \L^*_g(1+x) \right).
\end{equation}
The rate function $\L^*_g$ can be evaluated directly, with the result
that $\L_g^*(1+\e) \geq C\e^2$, where $C$ can be chosen
to be the constant $(6\ln 2)^{-1}$~\cite{BHLSW03}. Repeating the argument for
deviations below the mean and applying the union bound completes the
proof.
\end{proof}
As an aside, we note that the probability density function for
$\Tr(U_j \ph U_j^\dg P)$ was recently calculated exactly by Zyczkowski
and Sommers~\cite{ZS00}. In principle, this should allow for an exact
calculation of the rate function $\L_U^*$.
\begin{lemma}
  \label{lem:net}
  For $0<\e<1$ and $\dim{\cH}=d$ there exists a set ${\cal M}$ of
  pure states in $\cH$ with
  $|{\cal M}|\leq (5/\e)^{2d}$, such that
  for every pure state $\ket{\ph}\in\cH$ there exists 
  $\ket{\tilde{\ph}}\in{\cal M}$ with
  $\bigl\| \proj{\ph}-\proj{\tilde{\ph}} \bigr\|_1\leq\epsilon$.
  (We call such a set an \emph{$\epsilon$--net}.)
\end{lemma}
\begin{proof}
  We begin by relating the trace norm to the Hilbert space norm:
  \begin{equation}\begin{split}
    \bigl\| \ket{\tilde{\ph}}-\ket{\ph} \bigr\|_2^2
                    &=2-2{\rm Re}\,\bra{\tilde{\ph}}\ph\rangle  
\\
                    &\geq 1-|\braket{\tilde{\ph}}{\ph}|^2
\\
                    &= \left(\frac{1}{2}
                     \bigl\|\proj{\tilde{\ph}}-\proj{\ph}\bigr\|_1 \right)^2,
  \end{split}\end{equation}
  where the last line can be shown by evaluating the eigenvalues of 
  $\proj{\tilde{\ph}}-\proj{\ph}$. 
  Thus it will be sufficent to find an $\epsilon/2$--net for the Hilbert
  space norm. Let $\cM = \{\ket{\ph_i}:1\leq i \leq m\}$ be a maximal set
  of pure states satisfying $\| \ket{\ph_i} - \ket{\ph_j} \|_2 \geq \e/2$
  for all $i$ and $j$. (Such a set exists by Zorn's lemma.)
  By definition, $\cM$ is an $\e/2$--net for $\|\cdot\|_2$. We can
  then estimate $m$ by a volume argument. As subsets of $\RR^{2d}$,
  the open balls of radius $\e/4$ about each $\ket{\ph_i}$ are pairwise
  disjoint and all contained in the ball of radius $1+\e/4$ centered at
  the origin. Therefore,
  \begin{equation}
   m (\e/4)^{2d} \leq (1+\e/4)^{2d}.
  \end{equation}
\end{proof}

\begin{proof}\proofComment{Of theorem \ref{thm:bigRandom}}
Let $(U_j)_{j\geq 1}$ be i.i.d.~$\rU(d)$-valued random variables,
distributed according to the Haar measure. We will show that with high
probability the corresponding $R$ in Eq.~(\ref{eq:defr}) is
$\e$-randomizing.  The proof will consist of bounding
\begin{equation}
\Pr_U \left( \sup_\ph \left\|
    \frac{1}{n} \sum_{j=1}^n U_j \ph U_j^\dg - \frac{\id}{d}
    \right\|_\infty \geq \frac{\e}{d} \right)
= \Pr_U \left( \sup_\ph \sup_\ps \left|
    \frac{1}{n} \sum_{j=1}^n \Tr( U_j \ph U_j^\dg \ps ) - \frac{1}{d}
    \right| \geq \frac{\e}{d} \right).
\end{equation}
The optimizations over $\ph$ and $\ps$ can both be taken over pure states
only by the convexity of $|\cdot|$.
Fix a net of projectors $\cM = \{ X \}$ and let $\tilde{\ph}$ 
be the net point corresponding to $\ph$ so that 
\begin{equation}
\sup_\ph \| \ph - \tilde{\ph} \|_1 \leq \frac{\e}{2d}.
\end{equation}
Define $\tilde{\ps}$ similarly.  
Lemma \ref{lem:net} provides a net with
$|\cM| \leq \left( \frac{10d}{\e} \right)^{2d}$.
We can then proceed as follows:
\begin{eqnarray}
&\;& \Pr_U \left( \sup_\ph \sup_\ps \left|
    \frac{1}{n} \sum_{j=1}^n \Tr( U_j \ph U_j^\dg \ps ) - \frac{1}{d}
    \right| \geq \frac{\e}{d} \right) \nonumber \\
&\leq& \Pr_U \lpH  \sup_\ph \sup_\ps
    \frac{1}{n} \sum_{j=1}^n \left| \Tr( U_j \ph U_j^\dg \ps )
        - \Tr( U_j \tilde{\ph} U_j^\dg \tilde{\ps} ) \right| \\
&\;&  \hspace*{13ex}  
        + \llH \frac{1}{n} \sum_{j=1}^n \Tr( U_j \tilde{\ph} U_j^\dg 
        \tilde{\ps})  - \frac{1}{d} \rlH 
    \geq \frac{\e}{d} \rpH \nonumber \\
&\leq&  \Pr_U \left( \sup_\ph \sup_\ps \left|
    \frac{1}{n} \sum_{j=1}^n \Tr( U_j \tilde{\ph} U_j^\dg \tilde{\ps} )
        - \frac{1}{d}
    \right| \geq \frac{\e}{2d} \right) \label{eq:supsup}.
\end{eqnarray}
In the last inequality, we used the estimate
\begin{equation}
\left| \Tr \lpm U_j \ph U_j^\dg \ps \rpm
        - \Tr \lpm U_j \tilde{\ph} U_j^\dg \tilde{\ps} \rpm \right| 
\leq \big\| \ph - \tilde{\ph} \big\|_\infty + 
    \big\|  \ps -  \tilde{\ps} \big\|_\infty,
\end{equation}
which, because $\ph-\tilde{\ph}$ and $\ps-\tilde{\ps}$ are traceless 
and either zero or rank 2, is equal to 
$\smfrac{1}{2}\|\ph - \tilde{\ph}\|_1 + \smfrac{1}{2}\|\ps-\tilde{\ps}\|_1$.
This sum is then less than or equal to $\e / (2d)$ by 
construction of the net.

Next, we replace the optimization over the set of all pure states 
in Eq.~(\ref{eq:supsup}) by 
optimization over the net, use 
the union bound and apply lemma \ref{lem:conc}:
\begin{eqnarray}
&\;&  \Pr_U \left( 
    \underset{\mbox{{\tiny $\tilde{\ph},\tilde{\ps} \in \,$}}  \cM}{\max}
    \left| \frac{1}{n}
    \sum_{j=1}^n \Tr( U_j\tilde{\ph}  U_j^\dg \tilde{\ps} ) - \frac{1}{d}
    \right| \geq \frac{\e}{2d} \right) \\
&\leq& |\cM|^2 
    \underset{\mbox{{\tiny $\tilde{\ph},\tilde{\ps} \in \,$}} \cM}{\max} 
    \Pr_U \left( \left| \frac{1}{n}
    \sum_{j=1}^n \Tr( U_j \tilde{\ph} U_j^\dg \tilde{\ps} ) - \frac{1}{d}
    \right| \geq \frac{\e}{2d} \right) \\
&\leq& \left( \frac{10d}{\e} \right)^{4d}
    \exp\left( -\frac{C n\e^2}{4} \right),
\end{eqnarray}
where $C \geq (6 \ln 2)^{-1}$ is the same constant as in the proof of
lemma \ref{lem:conc}.  The existence of the desired $\e$-randomizing
map is guaranteed if the above probability is bounded away from $1$,
which is the case if $n > \smfrac{16d}{C \e^2} \log
(\smfrac{10d}{\e})$.  If $d > \smfrac{10}{\e}$, this is true when
when $n \geq 192 (\ln 2) \e^{-2} d \log d$.  

\end{proof}

\section{Randomization and the destruction of correlations}
\label{sec:characterize}
Our discussion in the previous section demonstrates that a quantum
operation $R$ constructed by averaging over $134 d (\log d) / \e^2$
randomly selected unitaries will be $\e$-randomizing
with high probability. In this section, our goal will be to
investigate the properties of general $\e$-randomizing maps so
the method used to construct $R$ will be immaterial as long as
\begin{equation}
\| R(\ph) - \smfrac{\id}{d} \|_\infty \leq \smfrac{\e}{d}
\end{equation}
for all $\ph$. The definition, it should be noticed, makes no mention
of the effect of $R$ on a system that is correlated with another 
system (we call this the ``environment'').  
Here we will analyze that effect, which will ultimately lead to a partial
characterization of all $\e$-randomizing maps. 

To start, it is 
easy to verify that an $\e$-randomizing $R$ properly destroys classical
correlations between the system being randomized and its environment:
\begin{lemma} \label{lem:destroyCorrelation}
Let $\r^{A\!B} = \sum_i p_i \; \ph_i^A \ox \ps_i^B$ be a separable state
and $R$ an $\e$-randomizing map on $A$. Then
\begin{equation}
\| (R\ox I)(\r^{A\!B}) - \smfrac{\id}{d} \ox \r^B \|_1
\leq \e.
\end{equation}
\end{lemma}
\begin{proof}
This is straightforward:
\begin{eqnarray}
\left\| (R\ox I)(\r^{A\!B}) - \smfrac{\id}{d} \ox \r^B \right\|_1
&=& \left\| \sum_i p_i [ R(\ph_i^A) \ox \ps_i^B -
    \smfrac{\id}{d} \ox \ps_i^B ] \right\|_1 \\
&\leq& \sum_i p_i  \left\| R(\ph_i^A) \ox \ps_i^B -
    \smfrac{\id}{d} \ox \ps_i^B \right\|_1 \\
&=& \sum_i p_i \| R(\ph_i^A) - \smfrac{\id}{d} \|_1 \\
&\leq& \e.
\end{eqnarray}
\end{proof}
Thus, for classically correlated states, approximate randomization
implies the destruction of correlations with other systems. Indeed,
for classically correlated states, finding an operation that will
destroy correlations is effectively the same thing as finding an
operation that will erase local information.

This equivalence fails dramatically for entangled states.  
(In contrast, any perfectly randomizing map {\em does}
destroy all possible correlations including entanglement.)  Indeed, if
we apply an $R$ constructed using the method of theorem
\ref{thm:bigRandom} to half of a maximally entangled state $\ket{\Ph}
= \smfrac{1}{\sqrt{d}} \sum_{j=1}^d |\ob{i}\>|\ob{i}\>$, then the
resulting state has rank at most $n = o(d^2)$, so
\begin{equation}
\| (R \ox I) (\Ph) - \smfrac{1}{d^2} \id \ox \id \|_1
\leq 2( 1 - n/d^2)
\xrightarrow{d \rar \infty} 2,
\end{equation}
meaning that $(R\ox I)(\Ph)$ and $\smfrac{1}{d^2}\id \ox \id$ can be
distinguished by a collective measurement with negligible probability
of error for large $d$. It goes without saying then that
the approximate randomization does not eliminate all the correlations
that were originally present in the maximally entangled state. Nonetheless,
the correlations that remain have been rendered invisible to local
operations and classical communication, recovering at least some of the
spirit of lemma \ref{lem:destroyCorrelation}.
\begin{lemma} \label{lem:LOCCinvisible}
Let $R$ be an $\e$-randomizing quantum operation,
$M = \{M_i\}$ be a positive operator-valued measure (POVM) 
that can be implemented using LOCC,
$p_i \!\! := \!\! \Tr(M_i (R \ox I)(\Ph))$ and 
$q_i \!\! := \!\! \Tr(M_i \smfrac{1}{d^2} \id)$.
Then $\| p - q \|_1 \leq \e$.
\end{lemma}
\begin{proof}
If $M$ can be implemented using LOCC then it will have the separable form
$M_i = X_i \ox Y_i$.
Moreover, we can assume without loss of generality that
$X_i$ and $Y_i$ are rank $1$ since refining the measurement can only
increase the $\ell_1$ distance between the outcome probability distributions.
Making use of the identities
$Y_i \Tr Y_i = Y_i^2$ and
$(I\ox Y_i)\Ph(I\ox Y_i) = \smfrac{1}{d} Y_i^T \ox Y_i$, the
demonstration is then direct:
\begin{eqnarray}
\| p - q \|_1
&=& \sum_i \llL \Tr[ (X_i \ox Y_i)(R \ox I) (\Ph) ]
    - \Tr[ (X_i \ox Y_i) \smfrac{\id}{d^2} ] \rlL \\
&=& \sum_i \Big| \Tr[ (X_i \ox \id)(R \ox I)
    (\smfrac{Y_i^T}{\Tr Y_i^T} \ox \smfrac{Y_i}{d})
    - \Tr[ (X_i \ox Y_i)\smfrac{\id}{d^2} ] \Big| \\
&=& \sum_i \Big| \Tr[ X_i( R(\smfrac{Y_i^T}{\Tr Y_i^T})
        - \smfrac{\id}{d} ) ] \Big|
    \Big| \Tr \smfrac{Y_i}{d} \Big| \\
&\leq& \sum_i \smfrac{\e}{d^2} \Tr(X_i)\Tr(Y_i) \leq \e,
\end{eqnarray}
where in the last step we have used that $\sum_i X_i \ox Y_i = \id$
and the fact that $R$ is $\e$-randomizing.
\end{proof}

A very similar proof demonstrates the more general
\begin{theorem} \label{thm:LOCCinvisible}
Let $M = \{M_i\}$ be a POVM that can be implemented using LOCC,
$p_i := \Tr(M_i \, (R \ox I)(\r^{AB}))$ and
$q_i := \Tr(M_i \, (\smfrac{1}{d} \id \ox \r^B))$. 
Then $\| p - q \|_1 \leq \e$.
\end{theorem}

Thus, while some kind of correlation can persevere when half of an
entangled state is randomized, that correlation will all be inaccessible
to LOCC measurements. In fact, the proof shows that it will be inaccessible
to all measurements that can be implemented using separable superoperators.
It's tempting to speculate that lemma \ref{lem:LOCCinvisible} provides
a characterization of all $\e$-randomizing operations. There is at least
a weak sense in which that is true: if the conclusion of lemma
\ref{lem:LOCCinvisible} holds for a map $R$, then for all states $\ph$,
\begin{equation} \label{eqn:traceRandom}
\| R(\ph) - \smfrac{\id}{d} \|_1 \leq \e.
\end{equation}
Recall that this condition is weaker than the operator norm definition
of $\e$-randomization that we use, however. This might suggest that
our definition is too strong and that this trace norm version might
be more easily characterized. Unfortunately, our proof of lemma
\ref{lem:LOCCinvisible} makes explicit use of the stronger condition.
We don't know if it would hold for $R$ only satisfying
Eq.~(\ref{eqn:traceRandom}).
The fact that $\e$-randomizing maps render the correlations of
entanglement invisible to LOCC also raises the question of their
relationship to the phenomenon of quantum nonlocality without
entanglement \cite{BDFMRSSW99}. The range of connections between
$\e$-randomizing maps and the physics of nonlocality will be further
developed in a upcoming paper~\cite{H03}.

\section{Quantum data hiding} \label{sec:dataHiding}

Theorem \ref{thm:LOCCinvisible} also immediately suggests another
application of $\e$-randomizing maps: quantum data hiding, the name
given to schemes for sharing bits or qubits between multiple parties
in such a way that the data cannot be accessed by LOCC operations
alone. We will focus on the bipartite case, where our methods provide
a protocol for hiding $l$ qubits in a bipartite state of roughly $2l$
qubits. This is far more efficient than previous constructions for
hiding either bits or qubits, where the best previous constructions
gave ratios that depended on the security requirements. To achieve $\d
= \e = 1/16$ in terms of the parameters introduced below, for example,
the separable Werner state construction of Ref.~\cite{EW02} requires
roughly $24$ qubits per hidden bit.  (Using, in the notation of that
paper, $K=4$ and $d=64$.)  To achieve qubit hiding, the construction
of Ref.~\cite{DHT02} would have multiplied that overhead by a further
dimension-dependent factor. 
(We note that by making additional assumptions about the operations
available to the parties,
it is possible to improve on the $1:2$ ratio between hidden and physical
qubits. It was recently discovered, for example, that in some systems 
with superselection rules, a ratio of $1:1$ is achievable~\cite{VC03}.)

\begin{definition}[Adapted from Ref.~\cite{DHT02}]
A \emph{$(\d,\e,p,q)$-qubit hiding scheme} consists of a CPTP encoding map
$E : \cB(\CC^p) \rar \cB(\CC^q)$ and a CPTP decoding map
$D : \cB(\CC^q) \rar \cB(\CC^p)$ such that
\begin{enumerate}
\item \emph{(Security)}
For all LOCC measurements $L$, as well as all states $\ph_0$ and $\ph_1$
on $\CC^p$,
\begin{equation}
\| L(E(\ph_0)) - L(E(\ph_1)) \|_1 \leq \e.
\end{equation}
\item \emph{(Correctness)} For all states $\ph$ on $\CC^p$,
    $\| (D \circ E)(\ph) - \ph \|_1 \leq \d$.
\end{enumerate}
\end{definition}
The security criterion obviously implicitly assumes some bipartite
structure $\CC^q \cong \cH_A \ox \cH_B$.  In our constructions, we
will set $\dim(\cH_A) = \dim(\cH_B) = d$.  Our main result is 
\begin{theorem}
For all $\d, \e > 0$ (satisfying  $\e^2 \log(40/\d^2)<1$) 
and sufficiently large $d$
$(d > \max\{\smfrac{36}{C\d^2},\sqrt{15/\e},21\})$, 
one can construct a $(\d,\e,p,d^2)$-qubit hiding
scheme, with
\begin{equation}
p = \frac{C \d^2 \e^2 d}{1188 \log d} \,.
\end{equation}
and $C = (6 \ln 2)^{-1}$.  
The encoding map is given by 
\begin{equation}
 R(\r) = \frac{1}{n} \sum_{j=1}^n U_j \rho U_j^\dg
\end{equation}
where $\{U_j\} \subset \rU(d^2)$, and $n=\smfrac{99}{C
\e^2} d \log d$. 
\end{theorem}
The main point, unfortunately obscured by the proliferation of constants
and conditions is simply this:
for these hiding schemes, the limiting ratio of hidden qubits to
physical qubits is
\begin{equation}
\lim_{d \rar \infty} \frac{\log p}{\log d^2} = \frac{1}{2}.
\end{equation}

The idea behind our approach is simple. We randomly choose the $U_j$
acting on the $AB$ system.  With $n$ sufficiently large, we can assure
that for all $\r$, $R(\r)$ is effectively indistinguishable (by LOCC
alone) from the maximally mixed state on $AB$.  Then we restrict the
input $\rho$ to have support on a sufficiently small subspace $S$ to
ensure that the subspaces $\{U_j S\}$ can be reliably distinguished by
a collective measurement on $AB$. This strategy is summarized in
figure \ref{fig:hiding}.  

\figureHiding 

We will prove the security and correctness in the next two subsections. 

\subsection{Security of the protocol}
The proof of security uses techniques similar to those in the proofs
of lemmas \ref{lem:conc} and \ref{lem:LOCCinvisible}, so we only give 
the outline here.  As usual, let $\{U_j: 1 \leq j \leq n \}$ be
$\rU(d^2)$-valued independent random variables distributed according
to the Haar measure.  For $0 < \e < 1$,
\begin{equation} \label{eqn:randSep}
\Pr_U \left( \sup_{\ph \in S} \sup_{\mbox{\tiny $X \! \ox \!Y$}} 
      \llH 
     \Tr \lbL \smfrac{1}{n} \sum_{j=1}^n (X \ox Y) U_j \,\ph\, U_j^\dg \rbL  
    - \smfrac{1}{d^2} \rlH \geq \frac{\e}{d^2} \right)
\leq 2 \, |\cM_d|^2 |\cM_p| \exp\left( - \frac{C n \e^2}{4} \right),
\end{equation}
where $X$ and $Y$ are $d$-dimensional rank $1$ projectors acting on
$A$ and $B$ respectively, and $\ph \in S$ is a $p$-dimensional pure
state (to be hidden).  $C$ is the same positive constant as in lemma
\ref{lem:conc}, $\cM_d$ and $\cM_p$ are $\smfrac{\e}{3d^2}$-nets for
$d$-dimensional and $p$-dimensional rank $1$ projectors respectively.
Eq.~(\ref{eqn:randSep}) can be proved in a way very similar to
lemma \ref{lem:conc}.
From lemma \ref{lem:net} we can choose 
$|\cM_d|^2 |\cM_p| = (15 d^2/\e)^{2(2d+p)}$. 
Whenever $d^2 > \max(15/\e,16)$ and $n \geq 33 (2d+p)
(\log d) / (C\e^2)$, the probability in
Eq.~(\ref{eqn:randSep}) is strictly less than $1/2$, in which case
more than half of the choices for $\{U_j\}$ are such that for all $\ph$,
$M$ and $N$, 
\bea
  \llH \Tr \lbL 
	\smfrac{1}{n} \sum_{j=1}^n (M \ox N) U_j \, \ph \, U_j^\dg \rbL  
    	- \smfrac{1}{d^2} \rlH \leq \frac{\e}{d^2}.  
\label{eq:goodhiding}
\eea

To finish the proof of security, we use arguments similar to those in
lemma \ref{lem:LOCCinvisible}.  Let $U_j$ be chosen so that $R(\ph) =
\smfrac{1}{n} \sum_{j=1}^n U_j \ph U_j^\dg$ satisfies
Eq.~(\ref{eq:goodhiding}).  Let $\{ X_i \ox Y_i \}$ be any POVM
implemented by LOCC, where $X_i$ and $Y_i$ are both rank $1$.  For any
state $\ph \in S$, let $p_i = \Tr( (X_i \ox Y_i) R(\ph))$ and $q_i =
\Tr((X_i \ox Y_i) \smfrac{\id}{d^2})$.  Using Eq.~(\ref{eq:goodhiding})
we find
\begin{eqnarray}
\| p - q \|_1
&=& \sum_i \left| \Tr( (X_i \ox Y_i) R(\ph) )
    - \smfrac{1}{d^2} \Tr(X_i \ox Y_i) \right| \\
&\leq& \sum_i \smfrac{\e}{d^2} \Tr(X_i \ox Y_i) \leq \e.
\end{eqnarray}
Therefore, when the conditions stated above on $n$, $d$ and $\e$ are
satisfied, the security condition is fulfilled with probability at least
$1/2$ for a random selection of unitaries.

\subsection{Correctness of the protocol}
To complete the construction of the data hiding scheme, we must also
show that the decoding can be performed reliably.  Let $S$ be a fixed
subspace of dimension $p$ in $\cH_A \ox \cH_B$ and let $P$ be the
projector onto $S$. Our decoding procedure will be given by the
\emph{transpose channel}~\cite{OhyaP93}, a generalization of the
pretty good measurement~\cite{HausladenW94}.  Specifically, let $N =
\sum_{i=1}^n U_i P U_i^\dg$ and $D_i = P U_i^\dg N^{-1/2}$. Our
decoding procedure $D$ will be given by performing the quantum
operation with Kraus elements $D_i$.  Our proof that this decoding
procedure works is via a reduction to the task of decoding classical
data, for which there are well-known criteria for the success of the
pretty good measurement~\cite{HausladenJSWW96}.

Fixing $0 < \a < 1$, our goal will be to ensure that
$|\<\ph|D_i U_i|\ph\>|^2 \geq 1-\alpha$ for all $i = 1,\ldots,n$ and
whenever $\ket{\ph}\in S$.  Then, $\<\ph| D
\circ R(\ph)| \ph\> \geq 1-\a$.  It would then follow by standard
inequalities~\cite{Fuchsv99} that
$\|D \circ R(\ph) - \ph \|_1 \leq 2 \sqrt{\a}$ and, choosing
$\a = \d^2 / 4$, that the correctness criterion is satisfied.
To begin, fix any basis $\cE = \{|\ob{j}\>\}$ for $S$.  Our strategy
is to first show that, for any given $i$ and $|\ob{j}\>$, the 
operation $D$ decodes $U_i\ket{j}$ correctly with high probability
by relating $D$ to the pretty good measurement for decoding classical
messages. We then show that $D$ succeeds on all pure input states 
by verifying that it succeeds simultaneously on a large enough set of bases
to effectively cover the set of all states.  

So, consider decoding the classical messages 
$i = 1,\cdots,n$ and $j = 1,\cdots,p$ by applying the pretty good
measurement to the set of states $\{ U_i \ket{\ob{j}}
\}_{ij}$.  The POVM elements are $M_{ij} = N^{-1/2} U_i \proj{\ob{j}}
U_i^\dg N^{-1/2}$.  This POVM can be implemented in two stages: first
$D$ is applied and the outcome $i$ is recorded, then the projective
measurement onto the basis $\cE$ is performed. To see this, observe
that
\begin{eqnarray}
\bra{\ob{j}} D_i \rho D_i^\dg \ket{\ob{j}}
&=& \bra{\ob{j}} P U_i^\dg N^{-1/2} \rho N^{-1/2} U_i P \ket{\ob{j}} \\
&=& \bra{\ob{j}} U_i^\dg N^{-1/2} \rho N^{-1/2} U_i \ket{\ob{j}} \\
&=& \Tr( \rho M_{ij} ).
\end{eqnarray}
In particular, this calculation also demonstrates that
\begin{equation}
|\bra{\ob{j}} D_i U_i \ket{\ob{j}}|^2 = \Tr(U_i\proj{\ob{j}}U_i^\dg M_{ij})
\end{equation}
so the decoding procedure $D$ succeeds on $\ket{\ob{j}}$ provided
that for each $i$, the pretty good measurement identifies
$U_i\ket{\ob{j}}$ with high probability. 
Applying the criterion of Hausladen \emph{et al.} for the success of
the pretty good measurement~\cite{HausladenJSWW96}, we find that
\begin{equation} \label{eqn:PGMcriterion} 1 - |\bra{\ob{j}} D_i U_i
\ket{\ob{j}}|^2 \leq \D_{ij} := \sum_{i' j' \neq i j} | \bra{\ob{j}}
U_{i}^\dg U_{i'} \ket{\ob{j'}} |^2.
\end{equation}
Notice that terms for which $j' \neq j$ and
$i' = i$ do not contribute to $\D_{ij}$
so its expectation value is
\begin{eqnarray}
\EE_U \D_{ij}
&=& \sum_{i' \neq i} \sum_{j'} \Tr[ \EE_U U_{i} \proj{\ob{j}} U_{i}^\dg
    U_{i'} \proj{\ob{j'}} U_{i'}^\dg ] \\
&=& \sum_{i' \neq i} \sum_{j'} \Tr[ \smfrac{\id}{d^2} \smfrac{\id}{d^2} ]
= \frac{(n-1)p}{d^2},
\end{eqnarray}
which is small provided $np \ll d^2$. We will be interested in
\begin{equation}
\Pr_U \left(
\D_{ij} \geq (1+\eta)\smfrac{(n-1)p}{d^2} \right),
\end{equation}
which by the left invariance of the Haar measure is equal to
\begin{equation}
\Pr_U \left(
\sum_{i' \neq i} \sum_{j'} |\bra{\ob{j}} U_{i'} \ket{\ob{j'}}|^2
\geq (1+\eta)\smfrac{(n-1)p}{d^2} \right).
\end{equation}
Invoking lemma \ref{lem:conc} with $\eta = 1/2$, the state
$\proj{\ob{j}}$ and projector $\sum_{j' = 1}^p \proj{\ob{j'}}$, the
probability that $\D_{ij}$ exceeds $\smfrac{3(n-1)p}{2d^2}$ is less
than or equal to $\exp( -C \, (n\!-\!1)\,p/4 )$ for the same positive
constant $C$ in lemma \ref{lem:conc}.  By the union bound, the
probability that this bad event happens for at least one of the
choices of $i$ 
is less than or equal to $n \exp( -C \, (n\!-\!1)\,p/4)$: 
\begin{equation}
\Pr_U \left(
\min_{i} |\<j|D_i U_i|j\>|^2 \leq 1- \smfrac{3 (n-1) p}{2 d^2} \right)
\leq n \exp(-C(n-1)p/4).
\label{eq:decode1}
\end{equation}

Now, as discussed earlier, our goal is to verify that $|\bra{\ph}D_i
U_i\ket{\ph}|^2$ will be large for all $\ket{\ph}\in S$.  Fix an
$\smfrac{\a}{2}$-net for pure states on $S$.  The size of this net
can be taken to be less than or equal to $(\smfrac{10}{\a})^{2p}$.
Extend each net point $\tilde{\ph}$ to a basis of $S$.  We have
\begin{eqnarray}
 \Pr_U \left( \min_{\tilde{\ph}} \min_{i} 
 |\<\tilde{\ph}|D_i U_i|\tilde{\ph}\>|^2 
 \leq 1 - \smfrac{3 (n-1) p}{2 d^2} \right)
&\leq& (\smfrac{10}{\a})^{2p} \, n \, \exp(-\smfrac{C}{4}(n\!-\!1)p)
\end{eqnarray}
by Eq.~(\ref{eq:decode1}) and the union bound.  The probability of
$D$ failing on at least one net point is less than $1/2$ if $n >
\smfrac{8}{C} \log(\smfrac{10}{\a}) + \smfrac{4}{C}
\smfrac{\log 2n}{p} + 1$.
Otherwise,
\begin{eqnarray}
|\<\ph|D_i U_i|\ph\>|^2 
&\geq& |\< \tilde{\ph} |D_i U_i|\tilde{\ph}\>|^2 - 
\llm |\<\ph|D_i U_i|\ph\>|^2 - |\<\tilde{\ph}|D_i U_i|\tilde{\ph}\>|^2 \rlm  
\nonumber 
\\
&\geq& 1 - \smfrac{3 (n-1) p}{2 d^2} - \smfrac{\a}{2} = 1 - \a 
\end{eqnarray}
for all $\ket{\ph}\in S$ by choosing $3 (n-1) p/ d^2 = \a$.  

That is essentially the end of the proof. All that remains is to make
appropriate choices for our various parameters.
Collecting all our requirements, we find that the correctness condition
$\|D \circ R(\ph) - \ph \|_1 \leq \d$ for all $\ph$ is satisfied with
probability at least $1/2$ if 
\begin{equation} \label{eqn:nCorrectness}
n >
\smfrac{8}{C} \log(\smfrac{10}{\a}) + \smfrac{4}{C}
\smfrac{\log 2n}{p} + 1,
\end{equation}  
$\a = \delta^2/4$
and $3(n-1) p/d^2 = \a$.  Recall that the security criterion is satisfied with
probability at least $1/2$ provided $n \geq
33 (2d+p) (\log d) / (C \e^2)$ and $d > \max(\sqrt{15/\e},4)$.
Restricting to $p \leq d$, we make the choice
\begin{equation}
n = \frac{99}{C \e^2} d \log d \,,~~~\mbox{and then~~~} 
p = \frac{C \e^2 \d^2}{1188} \frac{d}{\log d}.
\end{equation}
If, in addition, $d  > \max(\smfrac{36}{C\d^2}, 21)$ and
$\log(40/\d^2) < 1/\e^2$, a straightforward calculation shows that 
all our requirements are met.
Therefore, by the union bound, the probability that both the 
correctness and security criteria are satisfied is greater than $0$ for a 
random choice of $R$. As an example, when
$\e = \d = \smfrac{1}{16}$, and $d$ is sufficiently large, $\log p
\geq \log d - \log \log d - 30$.  Finally, both $\d$ and $\e$ can be chosen to
be any polynomial in $\smfrac{1}{\log d}$ without affecting the asymptotic
efficiency, and can be chosen to be $d^{-\a}$ for small $\a>0$ in order to
achieve security that is exponential in $2 \log d$, the number of physical
qubits, at the expense of a slightly reduced asymptotic 
efficiency $(1-4\a)/2$.

\section{Locking classical correlations} \label{sec:locking}
Define the maximum
classical mutual information that can be obtained by local measurements
$X \ox Y$ on a bipartite state $\r_{AB}$ as
\begin{equation}
I_c(\r) = \max_{X \ox Y} I(x:y),
\end{equation}
where $x$ and $y$ are random variables representing the outcomes of
measurements $X$ and $Y$ on $\r_{AB}$ and $I(x:y)$ is equal to
$H(x) + H(y) - H(x,y)$ for the Shannon entropy $H$. Now suppose that
$\r_{AB}'$ is obtained from $\r_{AB}$ by communicating $l$ classical bits
present in Alice's system to Bob. There are natural cryptographic
reasons to worry about the relationship between $I_c(\rho)$, $I_c(\rho')$
and $l$. Suppose, for example, that
an eavesdropper, initially uncorrelated with
a quantum state, can extract $I_c(\rho)$ bits of mutual information
about some secret classical data by performing a measurement on the state. 
If instead
the eavesdropper started with $l$ classical bits potentially correlated with
the secret, one would hope that the most the eavesdropper could learn
upon performing her measurement would be less than $I_c(\rho)+lD$ bits for
some constant $D$. The existence of locked classical correlations in the form
presented here demonstrates
that such bounds fail drastically in general. 
As a consequence then,
it is generally much more prudent to use the Holevo $\chi$ quantity
instead of the accessible information when bounding an eavesdropper's
information. ($\chi$ does obey simple bounds of the desired type.)

In their paper introducing the idea
of locked classical correlations, DiVincenzo \emph{et al.}~\cite{DHLST03} 
defined two figures of merit,
\begin{equation}
r_1 = \frac{I_c(\r)}{I_c(\r')}
\quad \mbox{and} \quad
r_2 = \frac{l}{I_c(\r') - I_c(\r)}.
\end{equation}
Ideally, the two should be small simultaneously: 
the first is the
ratio of the initial to the final information while the second measures
the ratio of the ``key length'' to the unlocked information. In their
paper, they found examples for which
$(r_1,r_2) \sim (\smfrac{1}{2},\smfrac{1}{\log d})$ and
$(r_1,r_2) \sim (\smfrac{1}{2\log d},\smfrac{1}{2})$. Here we show
that $r_1$ and $r_2$ can be made arbitrarily small at the same time,
meaning that the amount of information unlocked is large relative both
to the amount of information originally available \emph{and} relative
to the number of classical bits communicated from Alice to Bob.
\begin{theorem}
For all $\e,\d > 0$ there exist bipartite states $\r_{AB}$ with
$r_1 \leq \e$ and $r_2 \leq \d$. Alice's system may be taken to be
a classical system of $\log d + 3 \log \log d$ bits and Bob's a quantum
system of $\log d$ qubits provided $\e$ is smaller than some fixed constant,
$\log d > \smfrac{16}{C'' \e} \log \smfrac{20}{\e}$ (where
$C''$ is a positive constant) and
\begin{equation}
\d \geq \frac{3 \log \log d}{(1-\e/2)\log d}.
\end{equation}
\end{theorem}
As in the original work, the states we consider will have the form
\begin{equation}
\r_{AB} = \frac{1}{dn}
\sum_{i=1}^d \sum_{j=1}^n \proj{ij}_A \ox (U_j \proj{i} U_j^\dg)_B,
\end{equation}
where the $\{\ket{ij}_A\}$ and $\{\ket{i}_B\}$ are orthonormal,
$d = \dim(B)$ and the $U_j$ are unitary. Thus, $j$ can be thought of as
a label describing which orthonormal basis is used on Bob's system
to encode $i$. For such states, a
convexity argument (see Ref.~\cite{DHLST03}) quickly implies that 
\begin{equation} \label{eqn:lockingBound}
I_c(\r) \leq \log d + \max_\ph \frac{1}{n}
\sum_{ij} |\bra{\ph}U_j\ket{i}|^2 \log |\bra{\ph}U_j\ket{i}|^2.
\end{equation}
The communication of $j$, which requires $l=\log n$ bits, obviously yields
a state $\rho'$ for which $I_c(\r') = \log d + \log n$ so an investigation
of the locking properties of $\r$ will hinge on bounding the second
term of Eq.~(\ref{eqn:lockingBound}).
Letting $p_{ji} = |\bra{\ph}U_j\ket{i}|^2$ and
$p_j = (p_{j1},\ldots,p_{jd})$, this second term is equal to
$- \smfrac{1}{n} \sum_j H(p_j)$. (Note that $p_{ji}$ and $p_j$ are functions
of $\ph$, although we have suppressed the dependence in our notation.) 
As usual, we will proceed by selecting
the operators $U_j$ at random using the Haar measure, in which case this
average entropy quantity will be provably large. Indeed, a now familiar
type of calculation 
(see appendix \ref{sec:entropicUncertainty}, with
the substitution $\e \mapsto \e/2$)
demonstrates that there is a positive constant $C''$ such that
\bea \Pr \lpH \inf_\ph \frac{1}{n} \sum_{j=1}^n H(p_j) \leq
     (1\!-\!\e/2) \log d - 3 \rpH 
     \leq 
     (\smfrac{20}{\e})^{2d} 
	\exp \lpm \! -n \lpmm \smfrac{\e dC''}{4 (\log d)^2} - 1 \rpmm \rpm \,, 
	\label{eqn:entropicUncertainty}
\eea
provided $\e < 2/5$ and $d \geq 7$. 
Choosing $n=(\log d)^3$ and $\log d$ 
to be larger than 
$\smfrac{16}{C'' \e} \log \smfrac{20}{\e}$ then
ensures that the probability is bounded away from $1$. It's worth pausing 
to interpret this statement: it means that there is a choice of $n$ bases
that is highly incommensurate with all states $\ph$, in the sense that averaged
over bases, the entropy of the probability distribution induced by measuring
any fixed $\ph$ is almost maximal.
Returning to locking, we see that
there exists a choice of unitaries such that
\begin{equation}
I_c(\r) \leq \log d - [(1-\e/2)\log d - 3] = \smfrac{\e}{2} \log d + 3.
\end{equation}
We can then estimate, using the facts that $3/\log d < \e /2$ and $d \geq 7$,
\begin{eqnarray}
r_1 &\leq&
\frac{\smfrac{\e}{2} \log d + 3}{\log d + 3 \log \log d} \leq \e \\
r_2 &\leq&
\frac{3 \log \log d}{(1-\e/2)\log d}.
\end{eqnarray}

The general mathematical question we addressed in this section amounts to
quantifying the constraints imposed by entropic uncertainty
relations~\cite{D83,MU88} on typical observables, 
an interesting problem in its own right, 
regardless of its connection to locking classical correlations.
As such, and acknowledging that the approximations used here
were quite crude, it would be worth developing a more detailed 
understanding of the distribution of the quantity
$\min_\ph \sum_{j=1}^n H(p_j).$

\section{Discussion}
We have explored a range of cryptographic applications that are based on
concentration phenomena in high-dimensional inner product spaces.
Most results in quantum information theory exploit
regularities in the structure of the input or operations; Schumacher's
quantum noiseless coding theorem~\cite{S95,OhyaP93}, 
for example, is based on the fact
that for large $l$, a state $\rho^{\ox l}$ will be almost entirely supported
on an $l(S(\rho)+\d)$-qubit subspace. The results we presented
here are of a related but different character: the regularity we exploit
is inherent in the structure of $\CC^d$ and, therefore, doesn't require
any additional constraints.

Our first application was to demonstrate the existence 
of approximate private quantum channels
capable of achieving exponential security 
(as measured, for example, by an eavesdropper's accessible information)
in the number of encrypted qubits
while simultaneously using only about half as much key as the 
well-known perfectly secure constructions. The failure of bounds
on the size of the secret key from the exact case to apply in our 
approximate setting exposed a new distinction between quantum and 
classical correlations: classical correlations must be destroyed by
local randomization operations while quantum correlations can survive
such operations. 
Our second application built on this principle to find protocols for
LOCC hiding of bipartite quantum states capable of encoding roughly
$l$ qubits in $2l$ qubits, a significant 
improvement over previous constructions.
We ended by exhibiting states with locked classical correlations. Such
states can be used to perform surprising communication tasks but also
serve as a warning that accessible information is a potentially
volatile measure for use in security definitions. 

Our results here suggest a number of possible directions for future research.
One natural question is the optimality of the cryptographic protocols
we've described. While a simple rank argument ensures that 
the $1:1$ asymptotic ratio of secret key
bits to encoded qubits achieved by our approximate private quantum
channel is optimal assuming a unitary encryption map, there appear to
be technical obstacles to proving optimality in case of CPTP encryption
maps with unbounded output dimension.
(If the size of the encrypted state is a polynomial function of the
size of the message then the proof is straightforward.
One need only combine the argument of Ref.~\cite{DHT02}, Section IV with
the Fannes inequality~\cite{Fannes73}.)
Optimality of the $1:2$ ratio found for quantum data hiding represents
an even bigger challenge; we know of no convincing argument, beset by
technical obstacles or not.
At a finer level of detail, since our focus has been on asymptotic rates,
we haven't made any serious attempt to optimize the constants in our
constructions; it is likely that significant improvements and perhaps
simplifications could be found, particularly in the estimates leading to
the locking result. Finally, from a practical point of view, the most
pressing problem would be to find computationally efficient versions of
the constructions we have presented here. Our Appendix A provides one
step in this direction in the case of approximate private quantum channels:
instead of selecting unitary transformations from the full unitary group,
it suffices to select them at random from the set of products of 
Pauli operators. This random
selection is easily done in polynomial time and the Pauli operators are
easily implemented. Unfortunately, since the number of random selections
is exponential in the number of encrypted qubits, this simplification
does not yet yield a polynomial time construction.

\subsection*{Acknowledgments}
We thank Daniel Gottesman, Leonid Gurvits, Karol Zyczkowski and,
in particular, Charles Bennett for their helpful suggestions. 
PH and DL acknowledge the
support of the Sherman Fairchild Foundation, the Richard C. Tolman
Foundation, the Croucher Foundation  and the US National
Science Foundation under grant no. EIA-0086038. AW is supported by the
U.K. Engineering and Physical Sciences Research Council.

\appendix

\section{Beyond Haar measure} \label{sec:beyondHaar}
For some applications, a randomization procedure using a different
distribution over unitaries
would be preferable; in data hiding, for example, proving that the decoding
procedure succeeds with high probability would have been greatly 
facilitated by using Pauli operators instead of arbitrary unitaries.
The convenience appears to come
at a price, however. Rather than the strong operator norm estimate of 
section \ref{sec:Haar}, we have only been able to prove the corresponding 
result in trace norm for the general case. 

Throughout, we assume only that $\{ U_j: 1 \leq j \leq n\}$ are independent
$\rU (d)$-valued random variables with density $p(U)$ such that 
$\EE U_j \ph U_j^\dg = \id / d$. The density function $p(U)$, for example,
could consist of point masses concentrated at tensor products of Pauli
operators.

\begin{lemma}
For a fixed pure state $\ph$, $\EE[ \|
    \smfrac{1}{n} \sum_{j=1}^n U_j \ph U_j^\dg
    - \id / d \|_1 ] \leq \sqrt{d/n}$.
\end{lemma}
\begin{proof}
First we evaluate
\begin{equation}
\EE\left[ \left\|  \frac{1}{n} \sum_{j=1}^n U_j \ph U_j^\dg
    - \frac{\id}{d} \right\|_2^2 \right]
= \frac{1}{n^2} \sum_{ij} \EE \Tr( U_i \ph U_i^\dg U_j \ph U_j^\dg )
    - \frac{1}{d}.
\end{equation}
The expectation value is easy to calculate:
\begin{eqnarray}
\sum_{ij} \EE \Tr( U_i \ph U_i^\dg U_j \ph U_j^\dg )
&=& \sum_i \EE \Tr( U_i \ph U_i^\dg )
    + \sum_{i \neq j} \EE \Tr( U_i \ph U_i^\dg U_j \ph U_j^\dg ) \\
&=& n + \sum_{i \neq j} \Tr\left( \smfrac{\id}{d} \smfrac{\id}{d} \right) \\
&=& n + \smfrac{n(n-1)}{d}.
\end{eqnarray}
Therefore,
$\EE[ \smfrac{1}{n} \| \sum_{j=1}^n U_j \ph U_j^\dg
    - \frac{\id}{d} \|_2^2 ] = \smfrac{d-1}{nd}$.
By the Cauchy-Schwartz inequality and the concavity of the
 square-root function, we can then estimate
\begin{eqnarray}
\EE\left[ \left\| \frac{1}{n} \sum_{j=1}^n U_j \ph U_j^\dg
    - \frac{\id}{d} \right\|_1 \right]
&\leq&  \sqrt{d} \EE\left[ \left\| \frac{1}{n} \sum_{j=1}^n U_j \ph U_j^\dg
    - \frac{\id}{d} \right\|_2 \right] \\
&\leq& \sqrt{d} \left(
    \EE\left[ \left\| \frac{1}{n} \sum_{j=1}^n U_j \ph U_j^\dg
    - \frac{\id}{d} \right\|_2^2 \right] \right)^{1/2}.\
\end{eqnarray}
The lemma then follows by substitution and the trivial inequality $d-1 < d$.
\end{proof}
We'll also make use of Azuma's inequality:
\begin{lemma} \label{lem:martingale}
Let $(Y_j)_{j=1}^n$ be a sequence of real-valued random variables such that
$|Y_j| \leq 1$. Let
$S_n = \sum_{j=1}^n Y_j$ and $S_0 = 0$. If $\EE[Y_j|S_{j-1}]=0$, then
\begin{equation}
\Pr\left( \frac{1}{n} S_n \geq t \right)
\leq \exp\left( \frac{-nt^2}{2} \right).
\end{equation}
\end{lemma}
\begin{proof}
See, for example, Ref.~\cite{DZ93}.
\end{proof}
\begin{theorem} \label{thm:littleRandom}
For sufficiently large $d$ and $n = d \log d / \e^2$, there exists
a choice of $\{U_j\}_{j=1}^n$ in the support of $p$ such that the inequality
\begin{equation}
\left\| \frac{1}{n} \sum_{j=1}^n U_j \ph U_j^\dg - \frac{\id}{d} \right\|_1
\leq \e
\end{equation}
holds for all states $\ph$.
\end{theorem}
\begin{proof}
For $n,k \geq 1$, let $Z_n = \| \sum_{j=1}^n U_j \ph U_j^\dg - n\id / d \|_1$,
$S_k = \EE[ Z_n | U_1,\ldots,U_k ] - \EE[Z_n]$ and $Y_k = S_k - S_{k-1}$.
It's also convenient to introduce the notation $S_0 = Y_0 = 0$. Note that
\begin{eqnarray}
\EE[Y_k | S_{k-1} ]
&=& \EE[ \EE[ S_k | U_1,\ldots,U_{k-1} ] - S_{k-1} | S_{k-1} ] = 0.
\end{eqnarray}
Also, for fixed $(U_1,\ldots,U_n)$ and unitary $\hat{U}_k$, the triangle
inequality gives
\begin{eqnarray}
&\;& \left| \left\| \sum_{j=1}^n U_j \ph U_j^\dg - \frac{n\id}{d} \right\|_1
    - \left\| \sum_{j\neq k} U_j \ph U_j^\dg + \hat{U}_k \ph \hat{U}_k^\dg
    - \frac{n\id}{d} \right\|_1 \right| \\
&\leq& \| U_k \ph U_k^\dg - \hat{U}_k \ph \hat{U}_k^\dg \|_1 \leq 2
\end{eqnarray}
so $|Y_k| \leq 2$. An application of lemma \ref{lem:martingale} to
$S_n = \sum_{k=1}^n Y_k$ then tells us that
\begin{equation}
\Pr\left( Z_n - \EE[Z_n] \geq 2nt \right)
    \leq \exp\left(\frac{-nt^2}{2} \right).
\end{equation}
By the previous lemma, if $n \geq 4d / \d^2$, then $\EE[Z_n] \leq n\d/2$.
Therefore, when this condition holds, we find that
\begin{equation}
\Pr\left( Z_n  \geq n\d \right)
    \leq \exp\left(\frac{-n\d^2}{32} \right).
\end{equation}
Fix an $\e/2$-net $\cM$ with $|\cM|\leq (10/\e)^{2d}$. Then
\begin{equation}
\Pr\left( \sup_\ph \left\| \sum_{j=1}^n U_j \ph U_j^\dg 
	- \frac{n\id}{d} \right\|_1 \geq n \e \right)
\leq
\Pr\left( \underset{\ph \in M}{\max}
    \left\| \sum_{j=1}^n U_j \ph U_j^\dg - \frac{n\id}{d} \right\|_1
    \geq \frac{n\e}{2} \right).
\end{equation}
By the union bound and our previous calculations, this probability
is less than or equal to
\begin{equation}
\left(\frac{10}{\e}\right)^{2d} \exp\left(\frac{-n\e^2}{128}\right).
\end{equation}
If $n = d \log d / \e^2$, then this quantity goes to zero with increasing $d$.
\end{proof}

\section{Proof of Eq.~(\ref{eqn:entropicUncertainty})}
	\label{sec:entropicUncertainty}

Our goal is to prove that there is a positive constant $C''$ such that 
\bea \Pr \lpH \inf_\ph \frac{1}{n} \sum_{j=1}^n H(p_j) \leq
     (1\!-\!\e) \log d - 3 \rpH 
     \leq 
     (\smfrac{10}{\e})^{2d} 
	\exp \lpm \! -n \lpmm \smfrac{\e C'' d}{2(\log d)^2} 
	- 1 \rpmm \rpm  \,, 
\nonumber 
\eea
where $p_{ji} = |\<i|U_j^\dg|\ph\>|^2$.  

We begin by estimating the concentration of measure for the entropy of 
measurement of a random state.
Let $\ket{\ps}$ be a pure state chosen from the unitarily invariant measure
on $\CC^d$, $q_i = |\braket{i}{\ps}|^2$ and $f(\ket{\ps}) = H(q)$ the
Shannon entropy of the distribution $q$. We use a version of Levy's 
Lemma~\cite{MS86}:
\begin{lemma}[Levy] \label{lem:Levy}
Let $f : S^{k-1} \rar \RR$ be a function with Lipschitz constant $\s$. Then
\begin{equation}
\Pr\left( \big| f - \EE f \big| > \eta \right)
\leq 4 \exp\left( -C'k \eta^2 / \s^2 \right),
\end{equation}
for Haar measure on $S^{k-1}$ and $C' > 1/(220 \ln 2)$ a constant.
\end{lemma}
For our application, $k=2d$ and the Lipschitz constant can be expressed
in terms of the $q_i$:
\begin{eqnarray}
\s^2 = \sup_\ps \nabla f \cdot \nabla f 
&=& \frac{4}{(\ln 2)^2} \sum_{i=1}^d q_i (1 + \ln q_i)^2 \\
&\leq& \frac{4}{(\ln 2)^2} \sum_{i=1}^d q_i ( 1 + (\ln q_i)^2 ) \\
&\leq& \frac{4}{(\ln 2)^2} (1 + (\ln d)^2 ) \leq 8 (\log d)^2,
\end{eqnarray}
where the second inequality, true if $d \geq 3$, can be shown using Lagrange
multipliers. The 
expectation value of $f$ is $\log d - \D(d)$, with 
$\D(d) = \log d - 
  (\smfrac{1}{2} + \smfrac{1}{3} + \cdots \smfrac{1}{d})/(\ln 2)$, 
which converges to $(1-\g)/(\ln 2)$, where $\g$ is
Euler's constant (approximately $0.577$)~\cite{JRW94}. Using the 
estimate~\cite{Y91}
\begin{equation}
\frac{1}{2(d+1)} < \sum_{i=1}^d \frac{1}{i} - \ln d - \g < \frac{1}{2d},
\end{equation}
we can guarantee that $1/2 < \D(d) < 1$ if $d \geq 7$. Thus, 
choosing $\eta = 2 - \D(d)$ and setting $C'' = C' / 8$, we find
\begin{equation} \label{eq:eq46boundoneterm}
\Pr\left( H(q) < \log d - 2 \right) 
\leq 4 \exp\left( - \frac{dC''}{(\log d)^2} \right).
\end{equation}

We now move on to bounding $\sum_{j=1}^n H(p_j)$ for a given $\ph$.
This is easily done using the Chernoff bound~\cite{DZ93}:
if $X_1,\ldots,X_n$ are i.i.d. random variables such that $X_j \in [0,1]$
and $\EE X = \mu \geq \a \geq 0$, then
\bea
	\Pr \lpH \frac{1}{n} \sum_{j=1}^n X_j \leq \a \rpH 
	\leq \exp \lpm -n D(\a \| \mu) \rpm 
\eea
where $D(\cdot\|\cdot)$ is the binary divergence function 
\bea
 D(\a \| \mu) = \a \log \a - \a \log \mu 
              + (1-\a) \log (1-\a) - (1-\a) \log (1-\mu) \,. 
\eea
Let $X_j = 0$ whenever $H(p_j) < \log d - 2$ and $X_j = 1$ otherwise.
Then $(\log d - 2)X_j \leq H(p_j)$ and by Eq.~(\ref{eq:eq46boundoneterm}),
$\EE_U X_j \geq 1 - 4 \exp( -dC''/ (\log d)^2 )$. 
Choosing $\a = 1-\smfrac{\e}{2}$ and $\mu = \EE_U X_j$ 
in the Chernoff bound, we find
\begin{eqnarray}
\Pr\Big( \frac{1}{n} \sum_{j=1}^n H(p_j) \leq 
(1-\smfrac{\e}{2})(\log d - 2) \Big)
&\leq&
\exp \lpm 
 -n D \lpmm 1-\smfrac{\e}{2} \| 1 - 4 \exp( \smfrac{-dC''}{(\log d)^2} \rpmm \rpm.
\label{eq:singlestate}
\end{eqnarray}
The divergence can be bounded as follows:
\begin{eqnarray}
D\Big(1-\smfrac{\e}{2} \| 1 - 4 \exp( \smfrac{-dC''}{(\log d)^2})\Big)
&\geq& -H(\e/2) - \e + \frac{\e dC''}{2 (\log d)^2} \\
&\geq& \frac{\e dC''}{2 (\log d)^2} - 1.
\end{eqnarray}
The first inequality arises by neglecting the mixed term corresponding to
$\a \log \mu$ in the divergence, since it is 
always nonnegative. The second is valid whenever $\e < 1/5$, which
we assume from now on.

To extend Eq.~(\ref{eq:singlestate}) to all possible states, choose 
an $\smfrac{\e}{2}$-net $\cM$ for $d$-dimensional pure
states, with $|\cM| = (10/\e)^{2d}$.  Write $\tilde{\ph}$ for the
net point corresponding to $\ph$.  Let $p_j$ be as previously defined
in terms of $\ph$, and $\tilde{p_{j}}$ be similarly defined in terms
of $\tilde{\ph}$.  Using the union bound, 
\bea 
\Pr\Big( \min_{\tilde{\ph} \in \cM} \frac{1}{n} \sum_{j=1}^n H(\tilde{p_j})
\leq (1-\smfrac{\e}{2})(\log d - 2) \Big) \leq |\cM| 
\exp \Big[ -n \Big( \frac{\e dC''}{2 (\log d)^2} - 1 \Big) \Big]  \,. 
\eea
Furthermore, viewing $p_j$
and $\tilde{p_{j}}$ as postmeasurement states, the monotonicity of
the trace norm implies
\bea 
	\| p_j - \tilde{p_j} \|_1 \leq \smfrac{\e}{2}.
\eea
Then, applying Fannes' inequality~\cite{Fannes73} to the distributions
$p_j$ and $\tilde{p_j}$,
\bea
	| H(p_j) - H(\tilde{p_j}) | \leq \smfrac{\e}{2} \log d 
	-  \smfrac{\e}{2} \log \smfrac{\e}{2} 
	\leq \smfrac{\e}{2}\log d + 1.  
\eea
A substitution then completes the proof.

\bibliographystyle{unsrt}
\bibliography{rand}

\end{document}